\documentclass[11pt,a4paper]{article}
\usepackage{jheppub}
\usepackage{graphicx}
\usepackage[compat=1.0.0]{tikz-feynman}
\usepackage{amsmath}
\usepackage{bigints}
\usepackage[normalem]{ulem}
\usepackage{xcolor}
\usepackage{cancel}
\usepackage{mathtools}
\usepackage{verbatim}
\usepackage{slashed}
\usepackage{titlesec}
\title{Quantization of Interacting Galilean Field theories}
\author[a]{Kinjal Banerjee,}
\author[a]{Aditya Sharma}
\affiliation[a]{BITS-Pilani, KK Birla Goa Campus, NH 17B, Bypass Road, Zuarinagar,
Goa, India 403726}
\emailAdd{kinjalb@gmail.com}
\emailAdd{p20170442@goa.bits-pilani.ac.in, adityasharma.theory@gmail.com}
\abstract{We present the quantum field description of Galilean electrodynamics minimally coupled to massless Galilean fermion in $(3+1)$--dimensions. At classical level, the Lagrangian is obtained as a null reduction of a relativistic theory in one higher dimension. We use functional techniques to develop the quantum field description of the theory. Quantum corrections to the propagators and vertex are obtained upto first order and the theory is found to be renormalizable to this order. The beta function of the theory is found to grow linearly; the theory is not asymptotically free.}
\begin{document}
\maketitle

\section{Introduction}

Galilean symmetry is deeply rooted in the systems where the velocity involved is much smaller than the speed of light. It is characterized by a set of symmetry generators viz: spatial translations $(P_i)$, temporal translation $(H)$, homogeneous spatial rotations $(J_{ij})$, and Galilean boosts $(B_i)$. It is often encountered in condensed matter physics, non relativistic fluid dynamics and magnetohydrodynamics etc. (see \cite{Greiner2003}\cite{2013arXiv1306.0638T}\cite{Rousseaux_2005}\cite{2000physics..10042J}\cite{1998hep.th....9123J}). In addition to the isometries mentioned before, Galilean symmetry can be also conformally extended \cite{banerjee2019uniqueness}\cite{2018}\cite{2009} by the addition of conformal symmetry generators: dilatation $(D)$, spatial conformal transformations $(K_i)$ and temporal conformal transformations $(K)$. Galilean symmetry distinguishes from Lorentzian symmetry in terms of the scaling of space and time. Unlike Lorentzian physics, here space and time are kept on unequal footing and are scaled as
\begin{equation*}
\label{eqn:gallimit}
t \rightarrow t \;\;, \;\; x_i \rightarrow \epsilon x_i \;\;,\;\; \epsilon \rightarrow  0
\end{equation*}
The scaling of space and time (also called as Galilean limit) significantly modifies the expression for the symmetry generators of our spacetime. The Galilean conformal symmetry generators (\ref{eqn:galsym}-\ref{eqn:galcon}) can be obtained by taking the Galilean limit of Poincar\'{e} generators \cite{Bagchi:2014ysa}. The application of Galilean conformal symmetry in physics systems has been studied in \cite{2006c}\cite{2006d}\cite{Jackiw:2000tz}.\\ 

The Galilean conformal generators form a closed Lie algebra called finite Galilean conformal algebra (GCA)\cite{banerjee2019uniqueness}\cite{2018} (an overview of the conformal generators and GCA is discussed in section \ref{section:newton}). A remarkable feature of GCA is that it can be given an infinite lift to form an infinite dimensional Lie algebra (\ref{eqn:lift}), called full Galilean conformal algebra (fGCA). Recently, there has been an upsurge in constructing field theories consistent with fGCA \cite{banerjee2019uniqueness}\cite{2018}\cite{Bagchi:2014ysa}\cite{Bagchi:2015qcw}\cite{Bagchi_2020}\cite{2022arXiv220112629B}. The presence of infinite symmetries in such field theories at the classical level, raises an important question of whether or not the symmetries survive at quantum level. In order to answer this question, we first need to construct a quantum field description of Galilean field theory. Moreover understanding the quantization of interacting Galilean field theories is also important from condensed matter physics perspective where we encounter effective field theories with emergent gauge fields and hence interaction between gauge fields and matter fields such as fermion in a non-relativistic setting needs to be explored. This paper is a first step towards addressing this. In this paper, we present the 1 loop renormalization of interacting Galilean field theories consistent with fGCA in $(3+1)$--dimensions. Our interest lies particularly in quantizing Galilean electrodynamics coupled to massless Galilean fermion. \\

The case of Galilean electrodynamics (GED) is particularly interesting because it does not posses propagating degree of freedom\footnotemark \footnotetext{This was demonstrated first classically in \cite{banerjee2019uniqueness} by employing Dirac's constraint analysis and then later in quantum theory by observing the nature of pole in the propagator \cite{Chapman_2020} .}. Galilean electrodynamics constitutes two $U(1)$ gauge fields $A_t$ and $A_i$ that form a gauge multiplet $(A_t,A_i)$ along with an additional scalar field $\varphi$ \cite{banerjee2019uniqueness}\cite{Bergshoeff_2016} and it is closely related to some early works carried out on Galilean electromagnetism in 1973 in \cite{1973} (also see \cite{Rousseaux2013}\cite{Festuccia:2016caf}). It also emerges naturally in the study of non-relativistic string theory \cite{Ebert:2021mfu}\cite{Gomis:2020izd}\cite{Gomis:2020fui}\cite{2022arXiv220212698O}. In fact, it has been observed that a perturbation of a scalar field (in the adjoint representation $U(1)$) around a solitonic $D(d-2)$--brane reduces to Galilean electrodynamics \cite{2022arXiv220212698O}. Galilean fermions were first studied in \cite{Montigny2008PathintegralQO} and subsequently in \cite{2018}. Coupling Galilean fermion to Galilean electrodynamics introduces the propagating degrees of freedom in Galilean electrodynamics. \\

The complete quantum field description of a pure interacting Galilean field theory has not been attempted in the literature before. However, some recent work in this direction has already been initiated in \cite{Chapman_2020} where the renormalization of Galilean electrodynamics coupled to a Schr\"{o}dinger scalar in $(2+1)$--dimensions (called as scalar GED or sGED) has been discussed. The coupling of a Schr\"{o}dinger scalar to Galilean electrodynamics significantly modifies the renormalization group and leads to the family of non relativistic conformal fixed points. More recently, work on Galilean Yang-Mills in \cite{2022arXiv220112629B} has initiated the tree level quantization of Galilean field theories.\\

In this paper, we have constructed an interacting Galilean field theory by coupling Galilean electrodynamics to massless Galilean fermions (\ref{eqn:gqed}). At the classical level, the theory is obtained by `null reducing' Lorentzian electrodynamics coupled to Dirac Lagrangian, in one higher dimension. The method of null reduction is well known in the literature and has been applied to construct the Lagrangian of various non-relativistic field theories starting from a relativistic field theory in one higher dimension\cite{Bergshoeff_2016}\cite{Santos_2004}\cite{20.500.11850/488630}. To obtain the quantum field theory description we use path integral techniques. The first order quantum corrections to the theory are regularized using the cut-off regulators.  Owing to the unequal scaling of the space and time in Galilean field theories, we introduce the cut-off regulators $\Lambda$ and $\Omega$ in the momentum and energy sector respectively. The theory is found to be renormalizable and admits a strictly positive beta function which grows linearly with the running coupling. This behaviour is different to sGED in $(2+1)$--dimensions, where, the beta function for the coupling vanishes at all orders of the perturbation theory \cite{Chapman_2020}. The behaviour of the beta function is also different from its Lorentzian counterpart where it grows as a cube of the running coupling \cite{Peskin:1995ev}\cite{ryder_1996}. \\

This paper is organized as follows. In section \ref{section:newton}, we present a brief discussion on Galilean conformal symmetry and review some well-known results on GCA. In section \ref{section:null} we discuss the method of null reduction and use it to write down the Lagrangian for Galilean electrodynamics coupled to Galilean fermion. In section \ref{section:quantrenor}, we discuss the quantum ``properties" of the theory. We use the method of cut-off regularization to bring out the nature of divergences. We conclude the section by addressing the renormalization and asymptotic behaviour of the theory. In section \ref{section:discussion} we summarize our findings and discuss our future plans.

\section{Classical Aspects: A review}
\subsection{Galilean Symmetry: A Cursory Visit!}
\label{section:newton}

Galilean symmetry of a $(d+1)$-dimensional spacetime is characterised by symmetry generators viz- time translations $(H)$, space translations $(P_i)$, homogeneous rotations $(J_{ij})$ and Galilean boosts $(B_i)$. In an adapted coordinate system $x^\mu\equiv (t,x^i)$ we have
\begin{equation}
\label{eqn:galsym}
\begin{split}
&H=-\partial_t \;\;,\;\; P_i =\partial_i\;\;,\;\; J_{ij}= x_i \partial_j-x_j \partial_i\;\;,\;\; B_i= t \partial_i\\
\end{split}
\end{equation}
In addition to (\ref{eqn:galsym}), Galilean symmetry can be conformally extended by including the conformal symmetry generators: dilatation $(D)$ and spatial and temporal conformal transformations $(K_i, K)$ i.e
\begin{equation}
\label{eqn:galcon}
D=-(t \partial_t+ x^i\partial_i)\;\;,\;\;K_i= t^2\partial_i \;\;,\;\; K=-(t^2\partial_t+2x_i t \partial_i)
\end{equation}
We can write (\ref{eqn:galsym}) and (\ref{eqn:galcon}) excluding $J_{ij}$ in a condensed notation i.e
\begin{eqnarray*}
\label{eqn:algebra}
&L^n= -t^{n+1}\partial_t-(n+1)t^nx_i\partial_i \;\;\;\;\;\;\;\;\;\;(n=-1,0,1 \implies H,D,K)\\
&M_{i}^n=t^{n+1}\partial_i \;\;\;\;\;\;\;\;\;\;\;\;\;\;\;\;\;\;\;\;\;\;\;\;\;\;\;\;\;\;\;\;\;\;\;(n=-1,0,1 \implies P_i,B_i,K_i)
\end{eqnarray*}
The generators $L^n, M_{i}^n , J_{ij}$ forms the Lie algebra called as Galilean conformal algebra (GCA) given by
\begin{equation}
\label{eqn:lift}
\begin{array}{l}
\left[L^{n}, L^{m}\right]=(n-m) L^{n+m}\;\; ,\;\; \left[L^{n}, M_{i}^{m}\right]=(n-m) M_{i}^{n+m} \;\; , \;\; {\left[M_{i}^{n}, M_{j}^{m}\right]=0}\\                        
{\left[L^{n}, J_{i j}\right]=0 \;\;,\;\; \quad\left[J_{i j}, M_{k}^{n}\right]=M_{[j}^{n} \delta_{i] k} }
\end{array}
\end{equation}
A remarkable feature of GCA is that it closes for all integral values for $n$ and $m$. This gives GCA an infinite lift and the algebra is called full Galilean conformal algebra (fGCA). There also exists a geometrical way of arriving at the Galilean symmetry generators.  Galilean field theories live on a Newton-Cartan manifold and the Galilean conformal isometries are the subset of conformal isometries of a ``flat" Newton-Cartan spacetime (see \cite{2009}\cite{PhysRevD.31.1841}\cite{Duval_1993}\cite{2014}\cite{2021arXiv211213403L} for more details).\\ The symmetry generators $\{L^n, M^n_{i},J_{ij}\}$ can be used to study the behaviour of local fields under the fGCA. We employ scale-spin representation of fGCA (\ref{eqn:algebra}) to construct the action of symmetry generators on the fields \cite{2018}. If $\Phi$ is the most general field (i.e scalar or vector etc) then one can construct the following,
\begin{eqnarray}
\label{eqn:l}
&\delta_{L^n} \Phi&= \Big(t^{n+1}\partial_t +(n+1) t^n (x^l \partial_l+\Delta) \Big) \Phi-t^{n-1} n(n+1)x^k \delta_{B_k} \Phi \\[5pt]
\label{eqn:m}
&\delta_{M_{l}^n} \Phi&= -t^{n+1} \partial_l \Phi+(n+1)t^n \delta_{B_l} \Phi 
\end{eqnarray}
where $\Delta$ is the scaling dimension, $ \delta_{B_l} \Phi$ is the action of boost generator on the field $\Phi$ i.e for some $\Phi=\{ \varphi, A_i, \chi \} $  where $\varphi$ is a scalar field, $A_i$ is a vector field, $\chi$ is some spinor field. We shall put (\ref{eqn:l}) and (\ref{eqn:m}) to use in the next section to demonstrate the invariance of Lagrangian of Galilean spinor field under fGCA.

\subsection{Null Reduction: A Pedagogical Survey}
\label{section:null}
We shall now discuss null reduction in the context of field theories. The method of null reduction is an efficient and convenient way to construct a $(d+1)$-dimensional Galilean invariant theory from a $(d+2)$ dimensional relativistic field theory. It is widely discussed in the literature, see for example \cite{Bergshoeff_2016}\cite{Santos_2004}\cite{20.500.11850/488630}\cite{1995}. Let us briefly discuss the procedure and use it to construct a few nonrelativistic field theories.\\[5pt]
One begin with Minkowski metric in one higher spatial dimension. In an adaptive coordinate chart $x^{\tilde{\mu}}\equiv (x^0,x^i)$, where $i =1,2,3.....d+1$, we can write, 
\begin{equation}
\label{eqn:mink}
g=-dx^{0} \otimes dx^{0}+\delta_{ij}dx^i \otimes dx^j 
\end{equation}
Introduce the two null coordinates $t,s$ as
\begin{equation}
\label{eqn:tu}
s=\frac{1}{\sqrt2} (x^1 +x^0) \;\; ,\;\; t=\frac{1}{\sqrt2} (x^1 -x^0) 
\end{equation}
Using (\ref{eqn:tu}), one can cast (\ref{eqn:mink}) into
\begin{equation}
\label{eqn:Minknull}
g=ds\otimes dt+dt\otimes ds+\delta_{ij} dx^i \otimes dx^j = \eta_{\mu \nu} dx^\mu \otimes dx^\nu
\end{equation}
where $x^\mu =(s,t,x^i)$ and $\eta_{\mu \nu}$ is the Minkowski metric written in coordinate chart $x^\mu$. Clearly, both $\partial_s$ and $\partial_t$ generates the symmetry of Minkowski spacetime\footnotemark. \footnotetext{Infact, $(\partial_s, g)$ on a $(d+2)$ dimensional manifold is also called as flat Bargmann structure. The null reduction is often discussed in the context of the Bargmann structure; however, the Minkowski metric in a standard flat coordinate chart can always be cast into a Bargmann structure. See \cite{20.500.11850/488630}\cite{PhysRevD.31.1841}\cite{2014}  for more details.}The idea of null reduction is to ``reduce" the $(d+2)$ dimensional field theories along one of the null symmetry generators $\partial_s$ or $\partial_t$, hence the name null reduction. The resulting theory is a $(d+1)$ dimensional theory where the leftover null coordinate (either $s$ or $t$) acquires the status of non-relativistic time. An obvious question that comes to mind is why null reducing a $(d+2)$ dimensional spacetime leads to a $(d+1)$ dimensional non-relativistic spacetime. The reason is geometric\footnotemark. Factoring out a $(d+2)$ dimensional Minkowski spacetime by a null vector field $\partial_s$ (say) leads to a NC spacetime. For a detailed analysis, we request the reader to go through \cite{2009}\cite{20.500.11850/488630}\cite{PhysRevD.31.1841}\cite{2014}. From here onwards, we shall construct a few examples of non-relativistic field theories invariant under fGCA in $(3+1)$--dimensions.
\footnotetext{This can also be understood at the level of Lie algebra. Note that the Bargmann in $d$ dimensions is the sub algebra of Poincar\'{e} in $d+1$ dimensions, consisting of generators that commute with the null momentum, thus allowing us to construct a non relativistic theory stemming from a relativistic theory restricted along one of the null direction see for example \cite{20.500.11850/488630}\cite{2021arXiv211213403L}.}
\subsubsection*{Examples of Null reduction from field theory}
\textbf{Scalar field} \\[5pt]
Let us quickly apply the procedure of null reduction to the case of real scalar field theory since it is the simplest yet instructive example to start with. The Lagrangian $\mathcal{\tilde{L}}$, for a relativistic massless scalar field $\phi$, on a Minkowski spacetime is
\begin{equation}
\label{eqn:scalar}
\mathcal{\tilde{L}}=\frac{1}{2} \eta^{\mu \nu}(\partial_\mu \phi )(\partial_\nu \phi)
\end{equation}
where $\eta^{\mu \nu}$ is the metric tensor on the Minkwoski spacetime \eqref{eqn:Minknull}.
If $x^\mu= (t,s,x^i)$ are the coordinates then an obvious expansion of (\ref{eqn:scalar}) leads to 
\begin{equation}
\mathcal{\tilde{L}}= \frac{1}{2} \Big\{\eta^{ts} (\partial_t \phi) (\partial_s \phi) +\eta^{st}(\partial_s \phi) (\partial_t \phi) +\eta^{ij}(\partial_i \phi) (\partial_j \phi)\Big\}
\end{equation} 
To ``reduce" the Lagrangian along the null coordinate $s$, we impose that the scalar field $\phi$ is preserved along the symmetry generator $\partial_s$ i.e 
\begin{equation} 
\partial_s \phi=0
\end{equation}
This leads us to an expression for the Lagrangian $\mathcal{L}$, of a non-relativistic scalar field 
\begin{equation}
\label{eqn:nonrel}
\mathcal{L}_{scalar}= \frac{1}{2} \eta^{ij}(\partial_i \phi) (\partial_j \phi)
\end{equation}
Evidently, time derivatives have disappeared in (\ref{eqn:nonrel}). Thus, (\ref{eqn:nonrel}) does not exhibit any dynamics at all. The resultant equation of motion is just the Laplace equation. It would not harm us any to conclude that the non-relativistic real scalar field theory is an `uninteresting' field theory. In what follows, we shall try to construct some `interesting' field theories using null reduction. An immediate extension to the case of real scalar field is that of a complex scalar field. In fact, the null reduction of complex scalar field is more conventional since it leads to Schr\"{o}dinger action (see \cite{Bergshoeff_2016}\cite{20.500.11850/488630} for more details on the null reduction of complex scalar field). However, our major focus shall be to construct non relativistic spinor field theory and its interaction with Galilean Electrodynamics.\\[5pt]
\textbf{Electrodynamics}\\[5pt]
To begin with, we start with a relativistic theory which in this case happens to be Maxwell's Lagrangian $\mathcal{\tilde{L}}$ i.e, \begin{equation}
\mathcal{\tilde{L}}= -\frac{1}{4} \eta^{\mu \rho} \eta^{\nu \sigma} F_{\mu \nu} F_{\rho \sigma}
\end{equation}
where $F_{\mu \nu}=\partial_\mu A_\nu -\partial_\nu A_\mu $ and $A_{\mu}$ is the vector potential whose components in an adapted chart $x^\mu\equiv (s,t,x^i)$ on a $(d+2)$ dimensional spacetime are $A_{\mu}=(\varphi,A_t, A_i)$. Following along the lines of previous example, we ``reduce" the Lagrangian in the null direction $s$ by demanding
\begin{equation}
\partial_s \varphi= \partial_s A_t=\partial_s A_i=0
\end{equation}
The resultant Lagrangian $\mathcal{L}$ takes up the following form
\begin{equation}
\label{eqn:galED}
\mathcal{L}_{GED}= \frac{1}{2} \Big [ (\partial_t \varphi)^2 +2(\partial_j A_t)( \partial_j \varphi)-\frac{1}{2}F_{ij} F^{ij}-2(\partial_j \varphi )(\partial_t A_j ) \Big ]
\end{equation}
Note that the same Lagrangian was constructed in \cite{banerjee2019uniqueness} using Helmholtz conditions\footnotemark.\footnotetext{Helmholtz conditions are the necessary and sufficient conditions which when satisfied by a set of second-order partial differential equations, guarantees an \emph{action}. For more details on the method and its applications, see \cite{Bagchi_2020}\cite{10.2307/1989912}\cite{2016arXiv160201563N}\cite{PhysRevD.103.105001}.} The corresponding equation of motion are
\begin{eqnarray}
\label{eqn:eomged1}
&\partial_i\partial_i A_j-\partial_j\partial_i A_i+\partial_ j\partial_t \varphi=0\\
\label{eqn:eomged2}
&\partial_t\partial_t \varphi +\partial_i \partial_i A_t-\partial_i \partial_t A_i=0\\
\label{eqn:eomged3}
&\partial_i\partial_i \varphi=0
\end{eqnarray}

The invariance of the equations of motion and the Lagrangian under fGCA is demonstrated in \cite{banerjee2019uniqueness}.\\[5pt]
\textbf{Fermionic Field}\\[5pt]
\label{section:spinor}
This example is slightly more detailed because the results are not easily available in the literature. We shall begin with massless Dirac's Lagrangian in the coordinate chart
$x^{\tilde{\mu}}=(x^0, x^i)$ i.e, 
\begin{equation}
\mathcal{\tilde{L}}=i \bar{\psi}\gamma^{\tilde{\mu}} \partial_{\tilde{\mu}}\psi
\end{equation}
where $\psi =\begin{pmatrix}
\phi \\
\chi
\end{pmatrix}$ with $\phi$ and $\chi$ being a 2-component spinors themselves and $\gamma^{\tilde{\mu}}=(\gamma^0, \gamma^1, \gamma^i)$ (where $i=2,3,4$) are Dirac matrices. As before, we recast the Lagrangian into the double null coordinate chart $x^\mu\equiv(s,t,x^i)$ as
\begin{equation}
\label{eqn:dirac2}
\mathcal{\tilde{L}}=i \bar{\psi}\gamma^{s}\partial_s \psi+i \bar{\psi}\gamma^{t}\partial_t \psi+i \bar{\psi}\gamma^{i}\partial_i \psi
\end{equation}
For the present discussion,  the exact form of Dirac matrices in the coordinate chart $x^{\tilde{\mu}}$ is not required. However, one can always construct them in chart $x^{\tilde{\mu}}$ using \eqref{eqn:gamma}-\eqref{eqn:gamma1}.
The $\gamma$--matrices $\gamma^\mu=(\gamma^s, \gamma^t, \gamma^i)$ in the coordinate chart\footnotemark\;$x^\mu \equiv(t,s,x^i)$ are \cite{https://doi.org/10.1002/prop.2190371203}\cite{article}
\footnotetext{There exist a yet another irreducible representation of $\gamma$--matrices in the coordinate chart $x^\mu$, see \cite{1996} for more details. Note that our subsequent results on renormalization remains the same irrespective of the representation that we use.}
\begin{eqnarray}
\label{eqn:gamma}
&\gamma^s&= \begin{pmatrix}
0\;\;\;&\;\;\sqrt{2}\;I\; \\
0\;\;\;&\;\;\;\;0\\
\end{pmatrix}\;\; = \dfrac{1}{\sqrt{2}} (\gamma^0+\gamma^1)\\[5pt]
\label{eqn:gamma1}
&\gamma^t&=\begin{pmatrix}
0\;&\;\;0 \\
-\sqrt{2}\;I\;&\;\;0\\
\end{pmatrix} \;\; = \dfrac{1}{\sqrt{2}}(\gamma^1-\gamma^0)\\[5pt]
\label{eqn:gammai}
&\gamma^i&= \begin{pmatrix}
i\sigma^i\;\;\;\;&\;\;\;\;0\\
0\;\;\;&-i\sigma^i\\ 
\end{pmatrix} \;\; \text{where},\; i=2,3,4
\end{eqnarray}
where $\sigma^i$ are the usual Pauli matrices and $I$ is a $2 \times 2$ identity matrices. The $\gamma$ matrices (\ref{eqn:gamma}-\ref{eqn:gammai}) obey the standard Clifford algebra,
\begin{equation}
\label{eqn:clifford}
\{ \gamma^\mu, \gamma^\nu \}= -2\eta^{\mu \nu}
\end{equation}
where $\eta^{\mu \nu}$ is the metric tensor associated to the Minkowski line element (\ref{eqn:Minknull}) in the coordinate chart $\{x^\mu\}$.  The adjoint spinor $\bar{\psi}$ is defined as $\bar{\psi}=\psi^{\dagger} G$ where $G$ is the matrix that can be obtained from\footnotemark  \; \eqref{eqn:gamma} and (\ref{eqn:gamma1}). \footnotetext{Obviously, $G$ is same as $\gamma^0$, however in order to maintain the consistency with the notation of double null coordinate chart $x^\mu$, we choose to call $\gamma^0$ by $G$. Also, this is useful because we are interested in the null reduced version of the Dirac's Lagrangian. Owing to the fact that the null reduction leaves us with $\gamma^t$ and $\gamma^i$ as our Dirac matrices on the Newton-Cartan spacetime, it is therefore convenient to label $\gamma^0$ by $G$, thus allowing us to define the adjoint spinor $\bar{\psi}=\psi^\dagger G$.}
For our present case, we get
\begin{equation}
\label{eqn:g}
G=\begin{pmatrix}
0\;&\;I\\
I\;&\;0\\
\end{pmatrix} \equiv \frac{1}{\sqrt{2}}(\gamma^s-\gamma^t)
\end{equation}
We are now in a position to ``null reduce" (\ref{eqn:dirac2}). As before, we impose $\partial_s \psi=0$ to reduce  (\ref{eqn:dirac2}) to
\begin{equation}
\label{eqn:fermi}
\mathcal{L}_f=i\bar{\psi} (\gamma^t \partial_t+\gamma^i \partial_i)\psi
\end{equation}
where $\gamma^I=(\gamma^t,\gamma^i)$ satisfy the Clifford algebra on Newton-Cartan spacetime i.e,
\begin{equation}
\label{eqn:cliffNC}
\{\gamma^I, \gamma^J\}=-2 g^{IJ}
\end{equation}
where $g^{IJ}=diag(0,1,1,1)$ is the degenerate metric associated to the Newton-Cartan spacetime (see Appendix \ref{section:NC} for a brief discussion on Newton-Cartan spacetime or \cite{2009} for more details). Using (\ref{eqn:gamma}) and (\ref{eqn:g}), we can write the Lagrangian (\ref{eqn:fermi}) in two component notation as,
\begin{equation}
\label{eqn:lagspinor}
\mathcal{L}_f=-\sqrt{2}i \phi^\dagger \partial_t \phi -\chi^\dagger \sigma^i \partial_i \phi+\phi^\dagger \sigma^i\partial_i \chi
\end{equation}
The equations of motion corresponding to $\phi^\dagger$ and $\chi^\dagger$ are given by
\begin{equation}
\label{eqn:levy}
-\sqrt{2}i\partial_t \phi+\sigma^i\partial_i\chi=0\;\;,\;\; \sigma^i\partial_i \phi=0
\end{equation}
The invariance of the Lagrangian (\ref{eqn:lagspinor}) under fGCA is not manifest but can be checked by using (\ref{eqn:l}) and (\ref{eqn:m}). The action of Galilean conformal symmetry generators on the spinors $\phi$ and $\chi$ takes on the following form,
\begin{eqnarray}
\label{eqn:fgca1}
 &\delta_{L^n} \phi&= t^{n+1}\partial_t \phi+(n+1)t^n(x^l \partial_l \phi+\Delta \phi)\\[4pt]
  &\delta_{L^n} \chi&= t^{n+1}\partial_t \chi+(n+1)t^n(x^l \partial_l \chi+\Delta \chi)+\frac{i}{\sqrt{2}}n(n+1)t^{n-1}x^k\sigma_k \phi \\[4pt]
 &\delta_{M_{l}^n} \phi&= -t^{n+1} \partial_l \phi \\[4pt]
 \label{eqn:fgca2}
 &\delta_{M_{l}^{n}} \chi&=-t^{n+1}\partial_l \chi-\frac{i}{\sqrt2}(n+1)t^n \sigma_l \phi
 \end{eqnarray}
where $\Delta=3/2$ is the scaling dictated by the dynamics. The equation of motion are in agreement with the ones obtained in \cite{Montigny2008PathintegralQO}\cite{Levy-Leblond1967}. Note that we call $\psi$ and $\bar{\psi}$ to be Galilean fermions because they come from the null reduction of a relativistic fermionic theory (Dirac Lagrangian) in one higher dimension. Also the Lagrangian (\ref{eqn:lagspinor}) is invariant under fGCA (\ref{eqn:fgca1}-\ref{eqn:fgca2}). Note that the understanding of fermions in the context of Galilean field theories constructed from the intrinsic structures of Newton-Cartan spacetimes is lacking in literature and is an avenue of future work. Some of the work in this direction has already been initiated in \cite{Fuini:2015yva}.

Issues regarding the quantization (both canonical and path integral) of such theories have been addressed in \cite{Montigny2008PathintegralQO}. An interesting feature of theory (\ref{eqn:fermi}) is that it exhibits dynamics. We shall exploit this feature to couple it with Galilean electrodynamics, which shall render us an interacting Galilean invariant field theory. Since we are interested in its quantization, we shall name the theory Gaillean quantum electrodynamics (GQED).\\[7pt]
\textbf{Interacting Galilean Field Theory}\\[5pt]
\label{section:GalileanQED}
Having worked out the cases for Galilean electrodynamics and Galilean fermion, the stage is all set to discuss the possibility of their interaction, which also happens to be the theme of this paper. We shall first construct the Lagrangian of the theory and then carry out the quantum field analysis in the next section. As always we shall start with a relativistic theory and ``reduce" it to arrive at Galilean invariant field theory. For our present discussion, we start with relativistic Electrodynamics coupled with fermions. In a suitable coordinate chart $x^\mu \equiv (t,s,x^i)$, we begin with the following Lagrangian
\begin{equation}
\label{eqn:qed}
\mathcal{\tilde{L}}= -\frac{1}{4}F_{\mu \nu}F^{\mu \nu} +i\bar{\psi}\gamma^\mu D_{\mu}\psi
\end{equation}
where $F_{\mu\nu}=\partial_\mu A_\nu -\partial_\nu A_\mu$ is the field strength tensor, $D_\mu=\partial_\mu-igA_\mu$ is the covariant derivative, $\gamma^\mu$ are the Dirac matrices satisfying the Clifford algebra (\ref{eqn:clifford}) and $A_{\mu}=(\varphi, A_t, A_i)$ is the vector potential. The coupling is introduced via covariant derivative $D_\mu$. As before, we expand (\ref{eqn:qed}) and demand
\begin{eqnarray}
\partial_s \psi=\partial_s \varphi= \partial_s A_t=\partial_s A_i=0
\end{eqnarray}
along with (\ref{eqn:gamma}) to reduce it to 
\begin{eqnarray}
\label{eqn:gqed}
\mathcal{L}&=&\mathcal{L}_{GED}+\mathcal{L}_{f} +\mathcal{L}_{i}, ~ ~ ~ \text{where } \\
\mathcal{L}_{i}&=&g \Big[ \sqrt{2} \varphi \chi^\dagger \chi -\sqrt{2}A_t \phi^\dagger \phi +iA_{i}(\chi^\dagger \sigma^i \phi-\phi^\dagger \sigma^i \chi) \Big ]
\end{eqnarray}
and 
$\mathcal{L}_{GED}$ and $\mathcal{L}_f$ are given by (\ref{eqn:galED}) and (\ref{eqn:lagspinor}). For convenience of calculation, we borrow the $\gamma$-matrix notation defined in (\ref{eqn:gamma}) and rewrite $\mathcal{L}_i$ in four component notation as
\begin{equation}
\label{eqn:li}
\mathcal{L}_{i}=g \Big[ \varphi \bar{\psi} \gamma^s \psi+ A_t \bar{\psi} \gamma^t \psi + \bar{\psi} \gamma^i A_i \psi\Big ]
\end{equation}
and $\mathcal{L}_f$ is given by (\ref{eqn:fermi}). Since here are five gauge fields $(\varphi, A_t, A_i)$, in (\ref{eqn:li}) there are five terms in the interaction Lagrangian $\mathcal{L}_i$. 

The fermionic field $\psi$ carries a mass dimension $[\psi]=[\bar{\psi}]=3/2$ and the gauge field couplet $(\varphi, A_t,A_i)$ has a mass dimension of unity i.e, $[\varphi]=[A_t]=[A_i]=1$. Evidently, (\ref{eqn:gqed}) is an interacting field theory where the interaction is mediated by the coupling constant $g$.\\
The equation of motion for GQED obtained by varying (\ref{eqn:gqed}) with respect to the gauge fields $\{\varphi, A_t, A_i\}$ are\begin{eqnarray}
\label{eqn:varphi}
&\partial_t\partial_t \varphi +\partial_i \partial_i A_t-\partial_i \partial_t A_i-g\bar{\psi}\gamma^s \psi=0\\
\label{eqn:at}
&\partial_i\partial_i A_j-\partial_j\partial_i A_i+\partial_ j\partial_t \varphi-g\bar{\psi}\gamma^j \psi=0\\
\label{eqn:ai}
&\partial_i\partial_i \varphi-g\bar{\psi}\gamma^t \psi=0
\end{eqnarray}
and with respect to $\bar{\psi}$ reads
\begin{equation}
\label{eqn:psibar}
(\gamma^t D_t +\gamma^i D_i +ig\varphi \gamma^s) \psi=0
\end{equation}
Note that, unlike GED, we have a theory with propagating degree of freedom. The system of equations (\ref{eqn:varphi})--(\ref{eqn:psibar}) have been used to describe Land\'{e} factor of electron's magnetic moment (see \cite{article} for more details). The stage is all set to address the issue of its quantization, which we shall discuss next.

\section{Quantum field description of Galilean Quantum Electrodynamics}
\label{section:quantrenor}
In this section we address the quantization of Galilean quantum electrodynamics. To begin with, we write down its \emph{action} using the Lagrangian (\ref{eqn:gqed})
\begin{equation}
\begin{split}
S=\bigintsss dt d^3x \Big\{ \frac{1}{2} \Big [ (\partial_t \varphi)^2 +2(\partial_j A_t)( \partial_j \varphi)-\frac{1}{2}F_{ij} F^{ij}-2(\partial_j \varphi )(\partial_t A_j ) \Big ] \\
+ i \bar{\psi} \Big [\gamma^t \partial_t +\gamma^i \partial_i \Big ] \psi
+  g \Big[ \varphi \bar{\psi} \gamma^s \psi+ A_t \bar{\psi} \gamma^t \psi + \bar{\psi} \gamma^i A_i \psi\Big ]   \Big \} 
\end{split}
\end{equation}
Since Galilean quantum electrodynamics is a gauge theory, it is important to gauge fix the \emph{action}. Gauge theories as such generally do not admit a propagator unless we fix the gauge. We employ the Faddeev-Popov trick \cite{Peskin:1995ev}\cite{ryder_1996}\cite{FADDEEV196729}\cite{10.2307/j.ctv10crg0r} to write down the gauge fixed action\;i.e
\begin{equation}
\label{eqn:gaugefix}
\begin{split}
S=\bigintsss dt d^3x \Big\{ \frac{1}{2} \Big [ (\partial_t \varphi)^2 +2(\partial_j A_t)( \partial_j \varphi)-\frac{1}{2}F_{ij} F^{ij}-2(\partial_j \varphi )(\partial_t A_j ) \Big ] \\
-\frac{1}{2\xi} \;(\partial_t \varphi+\partial_i A_i)^2
+ i \bar{\psi} \Big [\gamma^t \partial_t +\gamma^i \partial_i \Big ] \psi
\\ +  g \Big[ \varphi \bar{\psi} \gamma^s \psi+ A_t \bar{\psi} \gamma^t \psi + \bar{\psi} \gamma^i A_i \psi\Big ]   \Big \} 
\end{split}
\end{equation}
where $\xi$ is the gauge fixing parameter. Note that the gauge fixing condition we have used to gauge fix\footnotemark\;the \emph{action} is $\Omega=(\partial_t \varphi+\partial_i A_i)$. 
\footnotetext{The same gauge fixed action can also be obtained by doing a null reduction of gauge fixed Lorentzian QED in one higher dimension.}
\subsection{Feynman Rules}
The inclusion of a gauge fixing term allows us to write down the propagators and the Feynman rules. For convenience, we introduce a vector $ \mathcal{X} ^I=(\varphi, A_t, A_i)$, the resulting propagators in \emph{momentum}\footnotemark\;space are (with $p=(\omega,p_i)\equiv (\omega,\bf{p})$ and $\bf{p}^2$ $= p_i p_i$ $\equiv \vec{p}^2$)
\begin{eqnarray}
\label{eqn:gaugeprop}
&\qquad \qquad \quad D_{IJ}(p)=
i\begin{bmatrix}
0\;\; &\;\; \dfrac{-1}{\vec{p}^2}\;\; &\;\; 0\\
\dfrac{-1}{\vec{p}^2}\;& \;\;\; (1-\xi)\dfrac{\omega^2}{\vec{p}^4}\;\; &\;\; -\dfrac{p_i}{\vec{p}^4} \omega (1-\xi)\\
0\;\; &\;\; -\dfrac{p_i}{\vec{p}^4} \omega (1-\xi)\;\;\;&\;\; -\dfrac{\delta_{ij}}{\vec{p}^2}+\dfrac{p_i p_j}{\vec{p}^4}(1-\xi)
\end{bmatrix}\\[10pt]
&G(p)\; = \;\Big< \bar{\psi} (p) \;\; \psi(p) \Big>\;\; = \;\Bigg ( \dfrac{i}{\gamma^t \omega +\gamma^i p_i} \Bigg ) 
\end{eqnarray}
\footnotetext{In Galilean physics, space and time are not on the same footing, hence the momentum and energy are to be treated as independent Fourier transform of space and time respectively.}
Also, the Feynman rule for the vertex is \\
\begin{equation}
\label{eqn:vertex}
\hspace{-2.5cm}V^I\;\;= \;\; i g\begin{bmatrix}
\gamma^s \\
\gamma^t \\
\gamma^i \\
\end{bmatrix} \;\; \equiv \;\; ig\gamma^I .
\end{equation}
We associate a diagram with gauge field propagator, fermion propagator and the vertex as follows,
\begin{itemize}
\item For fermion propagator ,\qquad \includegraphics[scale = 0.55]{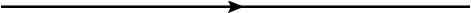}\;\;\; = \;\;$G(p)$
\item For gauge propagator ,\qquad \; $\it{I}\includegraphics[scale = 0.56]{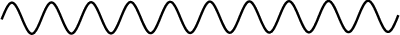}\it{J}\;\;\;= \;\; D_{IJ}(p)$
\begin{equation} 
\label{eqn:vert}
\end{equation}
\item \text{For each vertex} , \qquad \qquad \includegraphics[scale = 0.55]{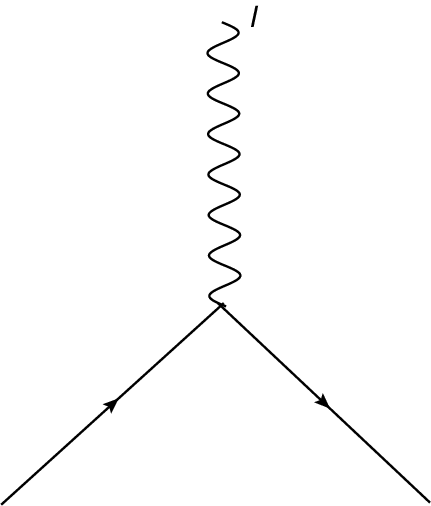}\qquad \;\; =\;\;$V^{I}$
\item An overall multiplicative factor of -1 for each internal fermion loop.
\end{itemize}
We note that, if we suppress the fermionic degrees of freedom, we end up with pure GED whose quantization is described by the propagator (\ref{eqn:gaugeprop}).  Some difference that can be drawn with Lorentzian QED are as follows:
\begin{itemize}
\item The expression for the propagator (\ref{eqn:gaugeprop}) indicates the absence of frequency ($\omega$) in the pole. This is suggestive of instantaneous transfer of information between two space-like separated points. 
\item The propagators $\big<\varphi, A_t\big>$ and $\big<A_t, A_i\big>$ are non zero. Physically, the non vanishing of the propagators means emergence of gauge field $A_t$ from a scalar field $\varphi$ and $A_i$ from $A_t$ respectively.
\end{itemize}
Similar features have been observed in sGED \cite{Chapman_2020}. However, such a behaviour is generally absent in Lorentzian field theories. We can quickly observe that the propagation $\big<A_t , A_i\big>$ can be turned off if we choose to work in the Feynman gauge (i.e, $\xi=1$), however, the propagator $\big<\varphi, A_t\big>$ can never be set to 0 by any choice of $\xi$.

\subsection{Regularization}
\label{section:regular}
To fully understand the features on Galilean quantum electrodynamics, we need to go beyond the tree level. We shall evaluate the 1 loop corrections to the fermion propagator, gauge field propagator and the vertex with a UV cut-offs $(\Omega, \Lambda)$  offered at large energy and momentum scales respectively. We begin our discussion by introducing the notion of superficial degree of divergence in context of Galilean field theories. The superficial degree of divergence in Galilean field theories is defined by the set $(\mathbb{D},\mathbb{F})$ where,
 \begin{eqnarray}
 \label{eqn:degreeD}
& \mathbb{D}&= 
\left(\begin{array}{c}
\text { Powers of $\omega$} \\
\text { in the numerator }
\end{array}\right)-\left(\begin{array}{c}
\text { Powers of $\omega$} \\
\text { in the denominator }
\end{array}\right)\\[7pt]
 \label{eqn:degreeF}
 &\mathbb{F}&=\left(\begin{array}{c}
\text { Powers of $\vec{p}$} \\
\text { in the numerator }
\end{array}\right)-\left(\begin{array}{c}
\text { Powers of $\vec{p}$} \\
\text { in the denominator }
\end{array}\right)
 \end{eqnarray}\\
Note that unlike Lorentzian field theories where the degree of divergence is characterised by a single number, here, in GQED we have a set $(\mathbb{D},\mathbb{F})$. This is because the unequal scaling of time and space forces one to consider two different cut-off scales viz. $\Omega$, and $\Lambda$.
Generally in gauge theories, the actual degree of divergence may differ from the one dictated by the superficial degree of divergence \cite{Peskin:1995ev}\cite{ryder_1996}. However, 
it is useful to have a naive idea about the extent to which the divergences can appear in a theory. We shall put (\ref{eqn:degreeD}) and (\ref{eqn:degreeF}) into use as and when required. We begin by evaluating corrections to the propagator and vertex. From here onwards, we shall work in the Feynman gauge i.e, $\xi=1$.

\subsubsection{Correction to the Fermion propagator}
\begin{figure}[h]
\centering
\includegraphics[scale=0.65]{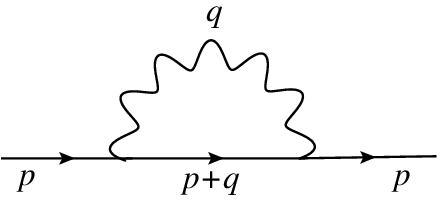}
\caption{Correction to the Fermion Propagator (self energy of a fermion)}
\label{fig:self1}
\end{figure}

The Feynman diagram for the same is drawn in Figure \ref{fig:self1}.
\\
The contribution of the correction term (self energy, $\Sigma$) can be obtained by integrating along an unconstrained variables $\vec{q}$ and $\omega_q$ i.e,
\begin{equation}
\label{eqn:correction}
\Sigma (p)= \dfrac{-g^2}{(2 \pi)^4} \bigintsss d \omega_q \; d^{3}\vec{q} \;\; \gamma^I D_{IJ}(\vec{q}) \gamma^J \bigg(\dfrac{i}{\gamma^t (\omega_q+\omega_p)+\gamma^i (q_i+p_i)}\bigg)
\end{equation}
where $\gamma^I$ is defined in equation (\ref{eqn:vertex}) and $D_{IJ}$ is the propagator (\ref{eqn:gaugeprop}).\\[5pt]
Simplifying (\ref{eqn:correction}) further, results in\footnotemark
\begin{equation}
\Sigma (p)= \frac{g^2}{(2 \pi)^4} \bigintsss d\omega_q \; d^3q \bigg(\frac{(\gamma^t (\omega_q+\omega_p)+\slashed{p}+\slashed{q})}{(\vec{q}+\vec{p})^2}\bigg)\frac{5 i^2}{\vec{q}^2}
\end{equation}\\
\footnotetext{From here onwards the contraction between $\gamma^i$ and any spatial vector quantity will be condensed into the slash notation (e.g $\slashed{a}=\gamma^i a_i$, for some $a_i$).}
The superficial degree of divergence takes the value $(\mathbb{D},\mathbb{F})=(2,0)$, which suggests quadratic and logarithmic UV divergences offered due to large energy and momentum respectively. In order to understand the nature of the divergence, we apply a UV cut-off $\Omega$ to the energy sector, the integral evaluates to 
\begin{equation}
\label{eqn:int}
\Sigma(p)=\frac{-5 g^2 }{8 \pi} \Bigg( \gamma^t \omega_p +\gamma^i p_i \Bigg)  \frac{\Omega}{|\vec{p}|}
\end{equation}
Note that, the integral does not admit any logarithmic divergence in the momentum sector (i.e, due to $\Lambda$) contrary to what the superficial degree of divergence $(\mathbb{F}=0)$ dictates. This is because the integral linear in $q$, vanish by anti-symmetricity of the integral. Also, the divergence in the energy sector is not quadratic as predicted by the $\mathbb{D}=2$. But as already pointed out, the superficial degree of divergence renders a naive idea about the divergences. Often, the actual divergence is softer in gauge theories than the superficial divergence. In this case however, the vanishing of the integral $ \int d \omega_q \; \gamma^t \omega_q$ restricts the divergence in the energy sector to the linear form.
Since, the cut-off $\Omega$,  has to be taken to the infinity, the correction term is linearly divergent. 
\subsubsection{Correction to the Gauge propagator}
\begin{figure}[h]
\centering
\includegraphics[scale=0.70]{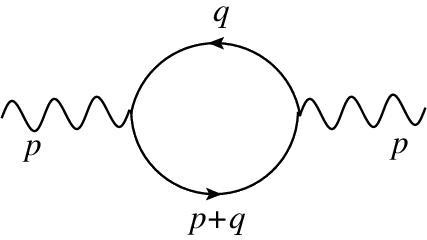}
\caption{Correction to the Gauge Propagator (self energy of the gauge field)}
\label{fig:self2}
\end{figure}

 The Feynman diagram is given in Figure \ref{fig:self2}.
\\
The contribution (self energy, $\Pi{}^{IJ}$) can be obtained by integrating the following integral
\begin{equation}
\Pi{ }^{IJ}(p) =\frac{-g^2}{(2 \pi)^4} \bigintsss d\omega_q \bigintsss d^3 q \; \text{Tr} \Bigg[ \gamma^I \gamma^J \Bigg(\frac{1}{\gamma^t (\omega_p+\omega_q)+\slashed{q}+\slashed{p}}\Bigg)\Bigg(\frac{1}{\gamma^t \omega_q+\slashed{q}}\Bigg)\Bigg]
\end{equation}\\
The superficial degree of freedom takes the value $(\mathbb{D},\mathbb{F})=(2,1)$, suggesting a quadratic divergence in the energy sector and a linear divergence in the momentum sector.  But it turns out, the diagram diverges linearly with a UV cut-off $\Omega$ i.e
\begin{equation}
\label{eqn:energygauge}
\Pi {}^{IJ}(p)= \dfrac{-5 g^2}{2 \pi^3 |\vec{p}^2|} \Lambda \Omega \; N^{IJ}
\end{equation}\\
where
\begin{equation}
N^{IJ}=\begin{bmatrix}
0\;\;&&1\;\;&&0\\[5pt]
1\;\;&&0\;\;&&0\\[5pt]
0\;\;&&0\;\;&&\delta_{ij}
\end{bmatrix}
\end{equation}\\[5pt]
As before, the integral is effective only upto a UV cut-off $\Omega$ and $\Lambda$. 

\subsubsection{Correction to the Vertex}
\begin{figure}[h!]
\centering
\includegraphics[scale=0.6]{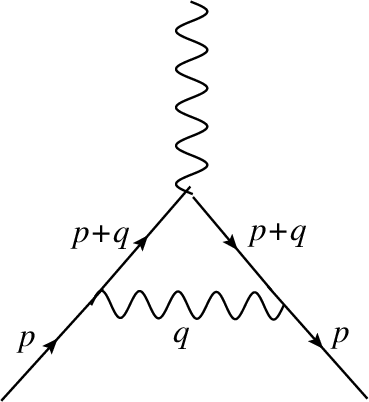}
\caption{Correction to the Vertex}
\label{fig:self3}
\end{figure}

The 1 loop correction to the vertex is given in Figure \ref{fig:self3}. 
The correction term ($\Gamma^I$) for the vertex evaluates to
\begin{equation}
\label{eqn:vertexcorr}
\Gamma^{I}(p)= \frac{5 g^3} {8 \pi} \frac{\gamma^I}{|\vec{p}|} \Omega
\end{equation}
As is obvious, in the limit $\Omega \to \infty$, the vertex correction term diverges linearly with a UV cut-off $\Omega$. This is not surprising since the superficial degree of divergence $\mathbb{D}=1$ suggests a linear divergence in the energy sector. In this case the correction term is convergent in the momentum sector which is in agreement with the superficial degree of divergence $\mathbb{F}=-1$. Clearly, all 1st order quantum corrections are divergent. Next, we shall discuss the renormalization of GQED at 1 loop.

\subsection{Renormalization}
Since, the 1 loop corrections depend upon UV cut-off scales, we need to renormalize the theory so that the physical quantities should not depend explicitly on the cut-offs. The mass dimension of  coupling constant $g$, turns out to be 0 which suggests that the theory might be renormalizable. To cancel the divergences in our theory, we shall introduce the counter-terms.
 We begin our discussion with the self energy of fermion.\\[5pt]
The self energy of fermion (\ref{eqn:int}) suggests that we must add a counter term 
\begin{equation}
\label{eqn:renorferm}
(\mathcal{L}_1)_{ct}=i B \bar{\psi}(\gamma^t \partial_t+\gamma^i \partial_i)\psi
\end{equation}
where B is chosen to render a finite fermion propagator upto order $g^2$ i.e,
\begin{equation*}
\includegraphics[scale=0.66]{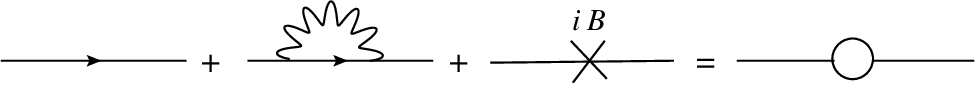}
\end{equation*}
Hence, one concludes that B must take the following value,
\begin{equation}
B=\frac{-5 g^2}{8 \pi |\vec{p}|} \Omega 
\end{equation} 
Defining $Z_2 =1+B$, the bare fermion wavefunction gets renormalized as
\begin{equation}
\psi_{(b)}=\sqrt{Z_2}\psi
\end{equation}
allowing us to write the bare Lagrangian for the fermionic part
\begin{equation}
(\mathcal{L}_1)_{(b)}=i\bar{\psi_{(b)}}(\gamma^t \partial_t+\gamma^i \partial_i)\psi_{(b)}
\end{equation}
We now turn to the gauge propagator. The self energy of gauge propagator (\ref{eqn:energygauge}) suggests a counter term is needed to remove the divergences i.e
\begin{equation*}
\includegraphics[scale=0.66]{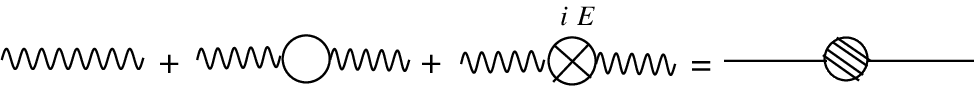}
\end{equation*}
The coefficient E is fixed to
\begin{equation}
E=\dfrac{-5 g^2}{2 \pi^3 |\vec{p}^2|} \Lambda \Omega
\end{equation}
Define $Z_3=1+E$, the bare gauge vector $\mathcal{X}_{I_{(b)}}$, gets renormalized as
\begin{equation}
\mathcal{X}_{I_{(b)}}=\sqrt{Z_3} \mathcal{X}_I
\end{equation}
giving us the bare Lagrangian for the gauge field
\begin{equation}
(\mathcal{L}_2)_{(b)}= \frac{1}{2}\Big [(\partial_t \varphi_{(b)})^2 +2(\partial_j A^t_{(b)})( \partial_j \varphi_{(b)})-\frac{1}{2}F_{ij}^{(b)} F_{(b)}^{ij}-2(\partial_j \varphi_{(b)})(\partial_t A^j_{(b)} ) \Big ] 
\end{equation}
Similarly, the following vertex renormalization,
\begin{equation*}
\includegraphics[scale=0.66]{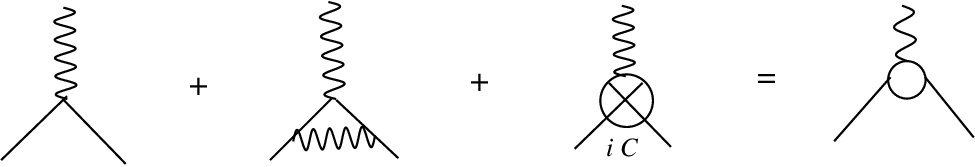}
\end{equation*}
 leads to the following counter term to be added to the Lagrangian i.e
\begin{equation}
(\mathcal{L}_3)_{ct} =C g \bar{\psi} \mathcal{X}_I\gamma^I \psi
\end{equation}
where $C$ takes the following value
\begin{equation}
C=\frac{-5 g^2}{8 \pi |\vec{p}|} \Omega 
\end{equation}
along with the following redefinition of the coupling
\begin{equation}
\label{eqn:gbare}
g_{(b)}=g \frac{Z_1}{Z_2 \sqrt{Z_3}}
\end{equation}
where, $Z_1= 1+C$ and $g_{(b)}$ is the bare coupling. The total Lagrangian in terms of bare quantities $(\mathcal{X}_{(b)}^I,\psi_{(b)},g_{(b)})$ is
\begin{equation}
\label{eqn:bareLag}
\begin{split}
\mathcal{L}_{(b)}= \frac{1}{2} \Big [(\partial_t \varphi_{(b)}^2 +2(\partial_j A^t_{(b)})( \partial_j \varphi_{(b)})-\frac{1}{2}F_{ij}^{(b)} F_{(b)}^{ij}-2(\partial_j \varphi_{(b)})(\partial_t A^j_{(b)} ) \Big ]\\
+i\bar{\psi_{(b)}}(\gamma^t \partial_t+\gamma^i \partial_i)\psi_{(b)}+ g_{(b)} \bar{\psi_{(b)}} \mathcal{X}_{I_{(b)}} \gamma^I \psi_{(b)}
\end{split}
\end{equation}
The fact that we have been able to absorb all the divergences into the definitions of bare quantities guarantees that GQED is renormalizable to this order. An important thing to note is that with the above choice of $B$ and $C$ we get $Z_1=Z_2$, allowing us to write,
\begin{equation}
\label{eqn:bareg}
g_{(b)}=\frac{g}{\sqrt{Z_3}}
\end{equation}
similar to what happens in Lorentzian QED \cite{Peskin:1995ev}\cite{ryder_1996}. The renormalized coupling $g$ is also called as running coupling.

\subsection{Asymptotic behaviour}
We now wish to understand how renormalized coupling $g$, changes with change in the cut-off. We note that the cut-offs $\Omega$ and $\Lambda$ can be algebraically related to each other by $\Omega=a \Lambda$, where $a$ is a dimensionless constant that captures the discrepancy in the two cut-offs  \cite{Chapman_2020}. This allows us to define\footnotemark\;the beta function $\beta(g)$ for our theory as
\begin{equation}
\beta(g)=\Lambda\frac{\partial g}{\partial \Lambda}=\frac{\partial g}{\partial \log{\Lambda}}
\end{equation}
\footnotetext{We could have defined the beta function in terms of $\Omega$ as $\beta(g)=\Omega\frac{\partial g}{\partial \Omega}$ and the results would not change.}
From (\ref{eqn:bareg}), we can deduce that\\
\begin{equation}
g_{(b)}=\dfrac{g}{\Big(1-\dfrac{5 g^2 a \Lambda^2}{2 \pi^3 |\vec{p}^2|}\Big)^{\frac{1}{2}}}
\end{equation}\\
By demanding the bare coupling $g_{(b)}$ to be independent of the cut-off, the beta function evaluates to
\begin{equation}
\label{eqn:beta}
\beta(g)=\Lambda\frac{\partial g}{\partial \Lambda}=2 g
\end{equation}
From (\ref{eqn:beta}), we can see that $\beta(g)$ is strictly positive ($\beta(g) >0$) and grows linearly with $g(\Lambda_0)$. The solution to (\ref{eqn:beta}) is
\begin{equation}
\label{eqn:sol}
g(\Lambda)=\frac{g(\Lambda_0) \Lambda^2}{\Lambda_0^2}
\end{equation}
where $\Lambda_0$ is the reference scale. We can deduce from (\ref{eqn:sol}) that at small momentum (or energy) range we have $\Lambda \sim \Lambda_0$ which implies $g(\Lambda)\sim g(\Lambda_0)$. However, at large values of $\Lambda$ i.e, $\Lambda >> \Lambda_0$, $g$ increases quadratically with $\Lambda$ i.e, the running coupling asymptotically gets stronger. Clearly, the theory becomes strongly coupled at large momentum. Note that, the behaviour of beta function (\ref{eqn:beta}) is different to what has been observed in sGED \cite{Chapman_2020} where the beta function for the coupling was shown to vanish at all orders of the perturbation theory. Unlike Lorentzian QED, this theory does not have a Landau singularity, since $\Lambda_0$ can never take the value zero. Note that we had started with a Galilean conformal field theory but the non vanishing of the beta function hints at the existence of global conformal anomalies in the theory as pointed out in \cite{2018}\cite{Jensen:2014hqa}\cite{Jain:2015jla}.
\section{Summary and Outlook}
\label{section:discussion}
Let us quickly summarize what we have accomplished in this paper. We have presented the quantum field description of Galilean electrodynamics coupled to massless Galilean fermions. The resulting theory is called Galilean quantum electrodynamics (GQED). The Lagrangian for the theory is obtained by employing the method of null reduction. The theory consists of a gauge field couplet $\mathcal{X}_I=(\varphi,A_t,A_i)$ of mass dimension 1, coupled to Galilean fermions $\psi$ of mass dimension $3/2$. The dimension of coupling constant $g$ between the gauge couplet and the fermion is 0. Note that, unlike Lorentzian QED, the gauge couplet consists of a spatial vector field $(A_i)$ and two scalar fields $(\varphi,A_t)$. We have gauge fixed the action by using Faddev-Popov method. Feynman rules are then obtained and the first order quantum correction are studied for propagators and vertices. Since this theory is not Lorentz invariant, we provide a prescription for quantization where we choose different cut-offs for momentum and energies. This leads us to the notion of superficial degree of divergence which is defined by a set $(\mathbb{D},\mathbb{F})$ in (\ref{eqn:degreeD})--(\ref{eqn:degreeF}). The divergences upto 1 loop are cured by cut-off regularization and renormalization. The beta function is obtained by algebraically relating the two cut-offs  and is found to grow linearly with the running coupling indicating the presence of a global conformal anomaly. 
\subsection*{Future Works}
Now that we have a viable route to quantize Galilean field theories, we plan to explore the renormalizability of GQED at all orders of the perturbation theory and explore the Ward identities. As shown in \cite{banerjee2019uniqueness}, classically GED was shown to be invariant under fGCA, which has infinite number of symmetry generators. It will be interesting to explore whether these infinite set of symmetries survive quantization or whether any of them become anomalous. Note that, the Galilean fermions constructed in this paper are via null reduction. The intrinsic geometric understanding of spinors in context of Newton-Cartan manifolds needs to be explored further. Some progress in this regard is made in \cite{Fuini:2015yva} where the procedure to construct covariant non relativistic Lagrangians for spinor fields in general Newton-Cartan background has been developed. 

Using the techniques developed in this paper we can also study the quantum field description of non abelian Galilean field theories (Galilean Yang-Mills,GYM) in the future. Some progress has already been made in \cite{Bagchi:2015qcw}\cite{2022arXiv220112629B}. Owing to the self-interaction in the theory, GYM already have propagating degrees of freedom. The issue of quantization of such theories at tree level has been addressed in \cite{2022arXiv220112629B}. 
We wish to study the renormalization of GYM with and without the matter degrees of freedom. Work in this direction is already in progress. Making a supersymmetric extension to Galilean field theories, both classically and then quantum mechanically, is another avenue of future work.

Also, many condensed matter systems can be found in the literature where Galilean symmetry is observed e.g \cite{Greiner2003}\cite{2013arXiv1306.0638T}\cite{2006d}. It will be interesting to see whether the quantum theory developed in this paper can be used to study such condensed matter systems.  Linking our theory to condensed matter applications would be one of the directions of the future research. 
\section*{Acknowledgement}
We would like to thank Arjun Bagchi, Rudranil Basu and Aditya Mehra for discussions, useful suggestions and for careful reading of the manuscript. We would also like to thank JaxoDraw \cite{BINOSI200476} for developing their free Java program for drawing Feynman diagram. KB would like to thank SERB CRG/2020/002035 for support. AS would like to thank the String theory group at IIT Kanpur, for useful comments and discussions.

\appendix

\section{Newton-Cartan Spacetime}
\label{section:NC}

A Newton-Cartan (NC) spacetime is a $(d+1)$ dimensional smooth manifold equipped with a degenerate contravariant metric $g^{IJ}$ along with a non-vanishing 1-form $\theta$ which also happens to be in the Kernel of $g$. \\[5pt]
For a flat NC spacetime $(\mathbb{R}\times \mathbb{R}^d)$, in an adaptive coordinate chart $x^I \equiv(t,x^i)$
\begin{equation*}
g=g^{IJ}\partial_I \otimes \partial_J \;\;,\;\; \theta=dt
\end{equation*}
The metric tensor $g$ takes the following form 
\begin{equation*}
g^{IJ}=
\begin{pmatrix}
0 & 0\\
0& I
\end{pmatrix} \equiv diag(0,I)
\end{equation*}
where $I$ is a $d \times d$ identity matrix. To establish the Clifford algebra on a Newton-Cartan spacetime, we use $\gamma^{I}=(\gamma^t, \gamma^i)$ where $\gamma^t$ and $\gamma^i$ are defined in (\ref{eqn:gamma}). Then it can be easily demonstrated that, 
\begin{equation*}
\{\gamma^I, \gamma^J\}=-2 g^{IJ}
\end{equation*}
is the degenerate Clifford algebra on a Newton-Cartan spacetime.

\bibliographystyle{unsrt}
\bibliography{Ref}

\end{document}